\documentclass[11pt,a4paper]{article}
\usepackage{preprint}

\hypersetup{
    pdftitle={MMS Observations of Kinetic Alfv{\'e}n Wave Turbulence and Steep Kinetic-Range Spectra in the Outer Plasma Sheet Boundary Layer},
    pdfauthor={Mani K Chettri, Rupak Mukherjee, Hemam D. Singh}
}

\title{MMS Observations of Kinetic Alfv{\'e}n Wave Turbulence and Steep
Kinetic-Range Spectra in the Outer Plasma Sheet Boundary Layer}

\author{%
Mani K Chettri\,\orcidlink{0009-0000-1368-9263}$^{1,*}$ \quad
Rupak Mukherjee\,\orcidlink{0000-0003-3955-7116}$^{1}$ \quad
Hemam D.\ Singh\,\orcidlink{0009-0003-3061-8944}$^{2,\dagger}$\\[0.5em]
\small $^{1}$Department of Physics, Sikkim University, Gangtok 737102, Sikkim, India\\
\small $^{2}$Department of Physics, Netaji Subhas University of Technology, New Delhi 110078, India\\[0.6em]
\small $^{*}$Email: \href{mailto:mkchettri8@gmail.com}{mkchettri8@gmail.com}\\[2pt]
\small $^{\dagger}$Corresponding author: \href{mailto:hemam.singh@nsut.ac.in}{hemam.singh@nsut.ac.in}%
}

\date{}

\begin{document}

\maketitle

\begin{abstract}
Energy dissipation and particle acceleration in the collisionless magnetotail plasma remain incompletely understood. While Kinetic Alfv{\'e}n Waves (KAWs) are widely hypothesized to mediate these processes, observational characterization of their spectral properties and dissipation signatures in magnetotail boundary layers remains limited. We report observations of KAW turbulence and parallel electric fields ($E_{\parallel}$) in the outer Plasma Sheet Boundary Layer (PSBL) using high-resolution burst-mode data from the Magnetospheric Multiscale (MMS) mission. For a crossing event on May 31, 2017, we identify broadband KAW turbulence characterized by a normalized electric-to-magnetic field ratio $\mathcal{R} = |\delta E_{\perp}|/(v_A|\delta B_{\perp}|) = 2.5 \pm 1.2$ exceeding the MHD limit, a spectral break near ion scales, a steep kinetic-range spectral slope ($\alpha = -3.48 \pm 0.13$), and low magnetic compressibility ($C_{\parallel} \approx 0.03$). We observe impulsive parallel electric field structures (up to 15~mV/m) and large-amplitude ion density fluctuations (up to 68\%) during intervals of enhanced wave activity. The steep spectral slope, steeper than theoretical predictions for undamped KAW cascades ($-7/3$ to $-8/3$), is consistent with substantial energy removal from the cascade at kinetic scales. The near-zero correlation between the KAW-band filtered (0.18--2.0~Hz) $E_{\parallel}$ waveform and ion density fluctuations (Pearson correlation coefficient, r $\approx 0.04$) suggests that the observed $E_{\parallel}$ structures are not straightforwardly organized by compressive density variations, consistent with dissipation through direct wave--particle interaction. Attribution to a specific damping channel (e.g., Landau damping) is not uniquely constrained by the present diagnostics. These observations support collisionless damping of KAW turbulence at kinetic scales in the intermediate-beta, outer PSBL of the terrestrial magnetotail.
\end{abstract}


\vspace{0.75em}

\section{Introduction}
\label{sec:introduction}

The Plasma Sheet Boundary Layer (PSBL) separates the cold, tenuous magnetotail lobes from the hot, dense central plasma sheet and is among the most dynamic regions in the terrestrial magnetosphere. Within the PSBL, the outer sub-region (lobe boundary layer) is defined by its proximity to the lobes: it is characterized by relatively low ion temperatures ($T_i \lesssim 1$~keV), moderate plasma beta ($\beta_i \sim 0.1$--$0.5$), and a mixture of lobe and plasma-sheet populations, distinct from the hotter inner PSBL adjacent to the central plasma sheet. Field-aligned currents, high-speed ion beams, and broadband electromagnetic fluctuations converge in this narrow transition zone, making the PSBL an important site for energy transport and conversion in the magnetotail system \citep{eastman1984, baumjohann1988}. Understanding the microphysics of this energy transfer remains a central challenge in magnetospheric physics.

Kinetic Alfv{\'e}n waves (KAWs) have emerged as a leading candidate for mediating energy transport and dissipation at kinetic scales where $k_{\perp}\rho_{i} \sim 1$ \citep{hasegawa1976particle, lysak1996}. Unlike ideal magnetohydrodynamic (MHD) Alfv{\'e}n waves, KAWs develop a finite parallel electric field component $E_{\parallel}$ through electron pressure gradients and finite ion gyroradius effects (and, where relevant, electron inertia). This $E_{\parallel}$ enables resonant energy exchange with field-aligned electrons, commonly discussed in terms of Landau damping, and can contribute to collisionless plasma heating and particle acceleration \citep{wygant2002, chaston2008}.

Finite-amplitude KAWs can drive compressive plasma responses through the ponderomotive force, generating density structures that locally modify the Alfv\'en speed \citep{hasegawa1976parametric, voitenko1998}. In magnetotail plasmas where $T_i/T_e \sim 1$--$3$, freely propagating ion-acoustic modes are strongly Landau-damped \citep{chen1984}, so compressive responses driven by KAW fluctuations may behave as quasi-modes: density perturbations continuously forced and rapidly dissipated \citep{terradas2022, verscharen2024}.

Testing whether the KAWs can couple to the localized density cavities or compressions generated by the ponderomotive force in the magnetospheric plasmas requires high-resolution observations capable of resolving the relationships between KAW fields and density perturbations. The presence of correlated density variations during intervals of KAW activity would support active nonlinear coupling, whereas the absence of this correlation would suggest that density variations are dominated by other boundary-layer processes. The Magnetospheric Multiscale (MMS) mission, with its high time resolution and four-spacecraft configuration, enables such measurements \citep{burch2016}. Recent MMS studies in the magnetosheath and magnetopause \citep{gershman2017, roberts2017} have reported KAW signatures exhibiting  nonlinear coupling of density variations with the magnetic field amplitudes of KAWs.  The effects of this coupling include the nonlinear generation of secondary electrostatic waves (like ion-acoustic and electron-acoustic waves), intense plasma heating, and the formation of solitary wave structures. \citet{ergun2015, stawarz2017} documented parallel electric fields reaching tens of mV/m during substorms and presented evidence consistent with electron Landau damping in turbulent plasmas \citep{chen2019}. In the magnetotail specifically, \citet{chaston2012} demonstrated energy transport by kinetic-scale electromagnetic waves in fast plasma sheet flows, and \citet{liu2022} reported KAW-associated energy dissipation inside magnetotail jets, establishing KAWs as an active dissipation mechanism in tail boundary regions.

The same MMS event on 31 May 2017 was previously analyzed by \citet{zhang2022observations}, who focused on identifying KAWs via the frequency-dependent electric-to-magnetic field ratio, characterizing the field-aligned Poynting flux, and providing direct evidence of electron acceleration through pitch-angle distributions and parallel potential-drop estimates. That study did not characterize the turbulence spectral properties across the ion-scale break, quantify magnetic compressibility as a mode diagnostic, analyze the statistical relationship between parallel electric field structures and density fluctuations, or evaluate cadence-matched energy conversion proxies as dissipation indicators. The present study addresses these open questions and makes the following distinct contributions relative to \citet{zhang2022observations}: (i) the measurement of the kinetic-range spectral slope ($\alpha = -3.48 \pm 0.13$) and its comparison with damped-KAW cascade theory; (ii) magnetic compressibility quantification ($C_\parallel \approx 0.03$) as an independent mode diagnostic; (iii) a KAW-band filtered $E_\parallel$--density correlation analysis; and (iv) cadence-matched $J_\parallel E_\parallel$ and $\mathbf{J}\cdot\mathbf{E}'$ energy-conversion proxies as dissipation indicators.

A key open question is whether KAW turbulence in the outer PSBL dissipates efficiently at kinetic scales and what spectral and energy-conversion signatures it imprints. In this paper, we address this question using MMS burst-mode data and the \textsc{PySPEDAS} analysis framework \citep{angelopoulos2019}, characterizing the kinetic-range spectral slope, magnetic compressibility, KAW-band $E_\parallel$--density coupling, and energy-conversion proxies for a PSBL crossing on May~31, 2017. Together, these diagnostics support efficient collisionless damping of KAW turbulence at kinetic scales in the outer PSBL.

\section{Instrumentation and Data Analysis}
\label{sec:methodology}

\subsection{MMS Instrumentation}

The MMS mission, launched by NASA in March 2015, comprises four identical spacecraft flying in a tetrahedral formation. This configuration enables simultaneous multipoint measurements of electromagnetic fields and plasma parameters at spatial separations down to electron kinetic scales. The MMS instrumentation suite provides the temporal resolution necessary to resolve kinetic-scale plasma dynamics \citep{burch2016}. Our study employs high-resolution burst-mode data from the MMS-1 spacecraft, selected for its optimal data coverage during the event interval (2017-05-31, 07:21:20--07:22:10~UTC). Magnetic field measurements derive from the Fluxgate Magnetometer (FGM), which samples the DC magnetic field at 128~vectors/s in burst mode \citep{russell2016}. This cadence resolves frequencies well above the local ion cyclotron frequency ($f_{ci} \approx 0.18$~Hz for the conditions reported here) while maintaining low noise floors essential for spectral analysis.

Electric field vectors are obtained from the Electric Double Probe (EDP) instrument suite, comprising the Spin-plane Double Probe (SDP) and Axial Double Probe (ADP) \citep{lindqvist2016, ergun2016}. In burst mode, EDP delivers three-dimensional electric field measurements at 8192~samples/s, enabling detection of high-frequency electrostatic fluctuations, solitary structures, and the parallel electric field signatures central to this investigation. Plasma moments (density, bulk velocity, and temperature for both ions and electrons) are provided by the Fast Plasma Investigation (FPI), which delivers full sky particle distributions at 150~ms cadence for ions and 30~ms for electrons in burst mode~\citep{pollock2016}.

\section{Plasma Environment and Event Overview}
\label{sec:observations}

On 2017-05-31, MMS-1 observed electromagnetic fluctuations in a boundary layer region of the magnetotail at GSE coordinates $(-20, -14, 4)~R_E$. The burst-mode data for this crossing provide a continuous 80-second record from 07:21:00--07:22:20~UTC; the overview time series (Figure~\ref{fig:overview}) is shown over the 07:21:15--07:22:10~UTC window, which captures the entry into and exit from the wave-active region. For quantitative spectral estimates, we restrict the analysis to a 50-second subset (07:21:20--07:22:10~UTC) during which background conditions and fluctuation statistics are approximately stationary, as required for PSD estimation and a single power-law fit. Table~\ref{tab:plasma_params} summarizes the plasma parameters measured during this analysis interval.

\begin{figure}
\centering
\includegraphics[width=0.85\linewidth]{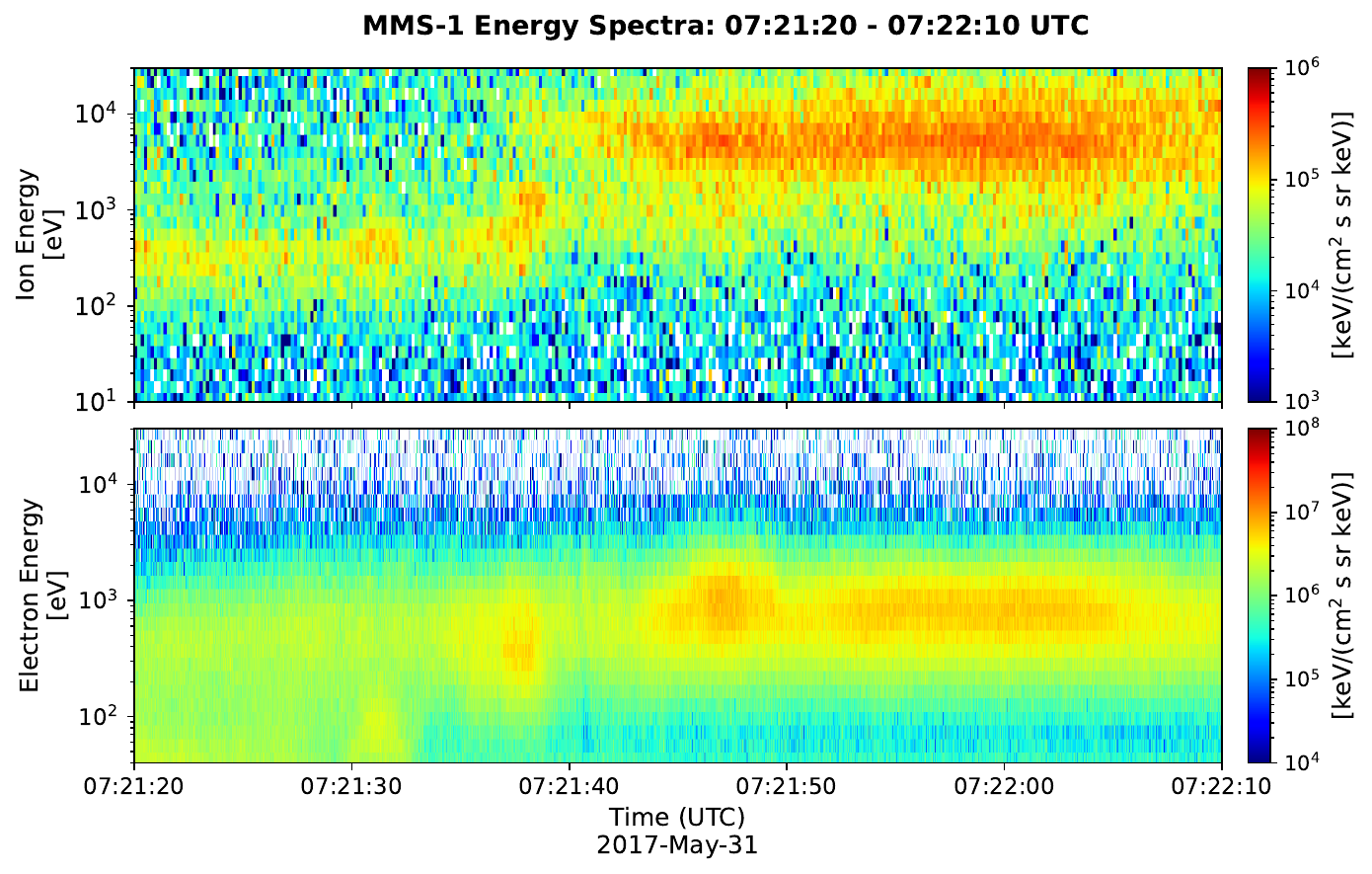}
\caption{Omnidirectional energy spectrograms for MMS-1 during the event interval. Top Panel: Ion energy spectrum showing a quasi-thermal population in the 0.2--5~keV range with a flux enhancement developing after 07:21:30~UTC. Bottom Panel: Electron energy spectrum showing flux variations in the 100~eV to 1~keV range. The temporal modulation of particle fluxes correlates with the interval of wave activity (07:21:20--07:22:10~UTC), consistent with local wave-particle interactions, though net energization requires detailed velocity distribution analysis.}
\label{fig:spectra}
\end{figure}

The environment is characterized by a moderate magnetic field ($B_0 = 11.5 \pm 0.5$~nT), low plasma density ($n_i = 0.15 \pm 0.03$~cm$^{-3}$), and magnetically dominated conditions ($\beta_i = 0.29$, $\beta_e = 0.12$). The relatively low ion and electron temperatures ($T_i = 0.65 \pm 0.55$~keV, $T_e = 0.27 \pm 0.10$~keV) indicate this crossing occurs in a cold transition region between the plasma sheet and tail lobes, consistent with the particle energy spectrograms in Figure~\ref{fig:spectra}. The large uncertainty, particularly in $T_i$, reflects significant temporal variability during the crossing. This ion temperature falls below the typical PSBL range of 1--5~keV, consistent with an outer-PSBL interval: the cold transition region between the hot central plasma sheet ($T_i \sim 3$--$7$~keV) and the tenuous tail lobes ($T_i < 0.1$~keV; \citep{baumjohann1988}). With an ion gyroradius $\rho_i \approx 320 \text{ km}$, inertial length $d_i \approx 595 \text{ km}$, and cyclotron frequency $f_{ci} \approx 0.18 \text{ Hz}$, the local plasma environment supports KAW dynamics. The electron beta satisfies $\beta_e > m_e/m_i \approx 5.4 \times 10^{-4}$, placing this plasma in the kinetic (rather than inertial) Alfv{\'e}n wave regime \citep{lysak1996}.

To visualize the particle environment hosting these fluctuations, Figure~\ref{fig:spectra} displays the omnidirectional differential energy flux for ions and electrons. The ion and electron temperatures reported above are obtained from MMS/FPI moment products and represent characteristic energies of the distributions, whereas the spectrogram shows how the measured flux is distributed across energy channels. It is evident from the ion spectrogram (top panel) that in the first half of the interval, the flux is intense at lower energies. Then, following some process, the flux-intense region moves to higher energies in the latter half with an enhancement. Here, we can not attribute this to a specific process, as net energization requires detail analysis of particle velocity distribution function. The electron spectrogram (bottom panel) shows flux variations in the 50 eV to 1 keV range. The temporal modulation of particle fluxes correlates with the interval of wave activity (07:21:20--07:22:10~UTC), suggesting local heating or acceleration through wave-particle interactions. To understand the specific energization mechanism, a detailed velocity distribution analysis is essential.

\begin{table}
\centering
\caption{Plasma parameters during the boundary layer crossing on 2017-05-31, 07:21:20--07:22:10~UTC. Values represent time-averaged means with standard deviations, obtained from MMS Level-2 data products (FGM magnetic field and FPI ion/electron moments) accessed via \textsc{PySPEDAS}; derived scales are computed from standard definitions.}
\label{tab:plasma_params}
\begin{tabular*}{\linewidth}{@{\extracolsep{\fill}}lcc@{}}
\toprule
Parameter & Symbol & Value \\
\midrule
\multicolumn{3}{l}{\textit{Observational parameters}} \\
Spacecraft location (GSE) & $(X, Y, Z)$ & $(-20, -14, 4)~R_E$ \\
Background magnetic field & $B_0$ & $11.5 \pm 0.5$~nT \\
Ion number density & $n_i$ & $0.15 \pm 0.03$~cm$^{-3}$ \\
Ion temperature$^{\dagger}$ & $T_i$ & $0.65 \pm 0.55$~keV \\
Electron temperature & $T_e$ & $0.27 \pm 0.10$~keV \\
Temperature ratio & $T_i/T_e$ & $2.37$ \\
Ion beta & $\beta_i$ & $0.29$ \\
Electron beta & $\beta_e$ & $0.12$ \\
Bulk ion velocity & $|\mathbf{V}_i|$ & ${\sim}150$~km/s \\
\midrule
\multicolumn{3}{l}{\textit{Derived characteristic scales}} \\
Alfv{\'e}n speed & $v_A$ & $657$~km/s \\
Ion gyroradius & $\rho_i$ & $320$~km \\
Ion inertial length & $d_i$ & $595$~km \\
Ion cyclotron frequency & $f_{ci}$ & $0.18$~Hz \\
\midrule
\multicolumn{3}{l}{\textit{Wave properties}} \\
RMS perpendicular electric field & $\langle \delta E_{\perp}^2 \rangle^{1/2}$ & $1.00$~mV/m \\
RMS perpendicular magnetic field & $\langle \delta B_{\perp}^2 \rangle^{1/2}$ & $0.39$~nT \\
\bottomrule
\multicolumn{3}{l}{$^{\dagger}$Large uncertainty reflects significant temporal variability during the crossing.}
\end{tabular*}
\end{table}

\section{Results}
\label{sec:results}

\subsection{KAW Identification}

\begin{figure}[htbp]
\centering
\includegraphics[width=0.85\linewidth]{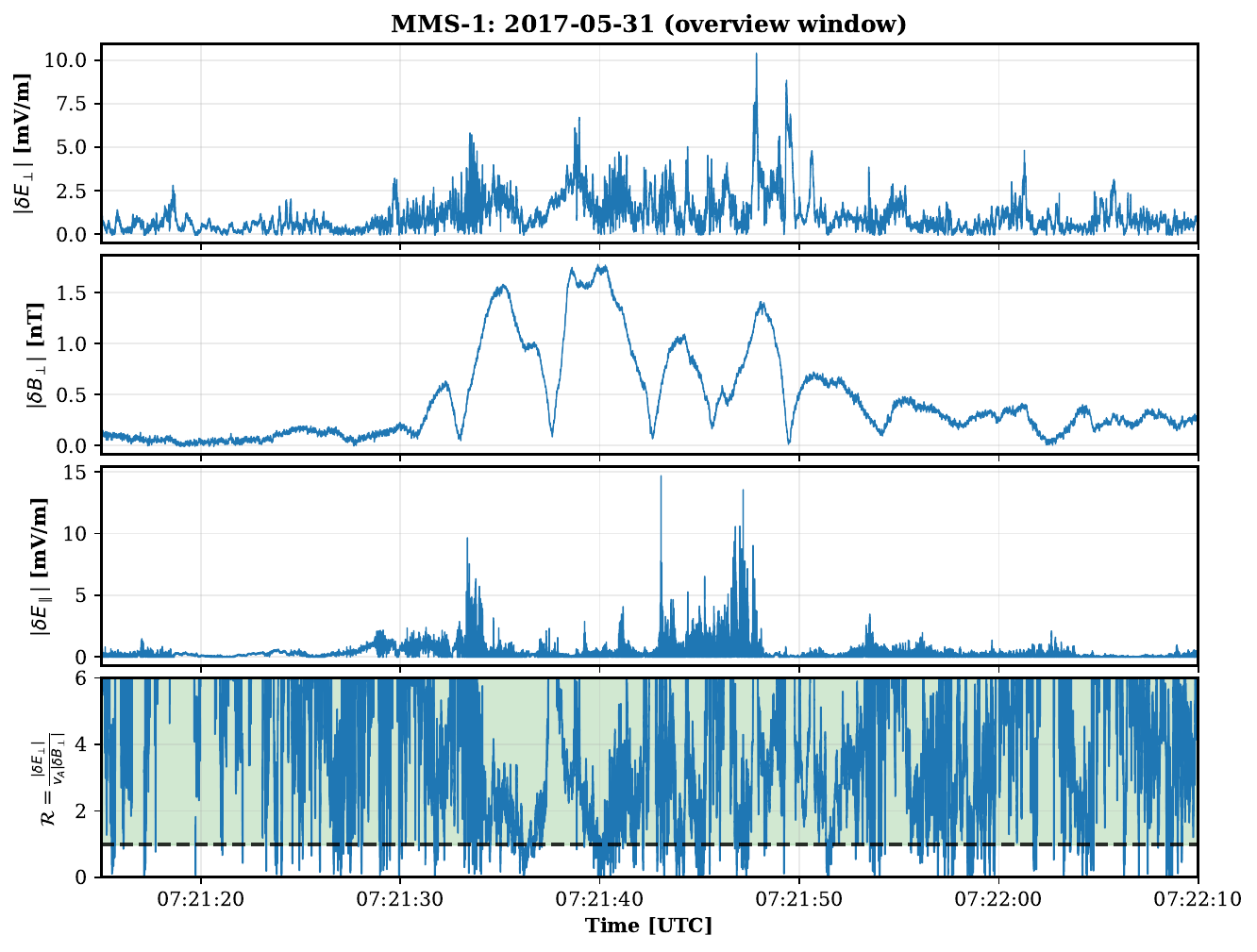}
\caption{Overview of MMS-1 burst-mode observations during a PSBL crossing on 2017-05-31 (07:21:15--07:22:10~UTC). Panels show: (a) Perpendicular electric field $|\delta E_{\perp}|$; (b) Perpendicular magnetic field $|\delta B_{\perp}|$; (c) Parallel electric field $|\delta E_{\parallel}|$ showing impulsive structures up to 13--15~mV/m;  (d) Normalized ratio $|\delta E_{\perp}|/(v_A|\delta B_{\perp}|)$ (blue) with mean value $2.5 \pm 1.2$, consistently exceeding the MHD limit (dashed line at unity), characteristic of KAW turbulence.}
\label{fig:overview}
\end{figure}

Figure~\ref{fig:overview} presents detailed burst-mode observations of the electromagnetic fluctuations over the observation window 07:21:15--07:22:10~UTC. Panels (a) and (b) show intense, broadband fluctuations in the perpendicular electric ($|\delta E_{\perp}|$) and magnetic ($|\delta B_{\perp}|$) fields, with root-mean-square amplitudes consistent with the values in Table~\ref{tab:plasma_params}. Panel (c) reveals impulsive parallel electric field structures ($|\delta E_{\parallel}|$) with peak amplitudes reaching $\sim$13--15~mV/m. Such elevated parallel electric field amplitudes are comparable to those observed during active reconnection events, double-layer structures, and intense wave-particle interaction regions in the magnetotail \citep{ergun2015, stawarz2017}.

A key diagnostic for distinguishing KAWs from ideal MHD Alfv{\'e}n waves is the ratio of perpendicular electric to magnetic field fluctuations normalized by the Alfv{\'e}n speed:
\begin{equation}
\mathcal{R} = \frac{|\delta E_{\perp}|}{v_A |\delta B_{\perp}|}.
\label{eq:ratio}
\end{equation}
For ideal MHD Alfv{\'e}n waves, $\mathcal{R} = 1$, whereas KAWs exhibit $\mathcal{R} > 1$ due to the enhanced electric field arising from finite ion gyroradius effects at $k_{\perp}\rho_i \sim 1$ \citep{stasiewicz2000, chaston2008}. The field amplitudes $|\delta E_\perp|$ and $|\delta B_\perp|$ are evaluated from measurements bandpass-filtered to the KAW frequency range ($f_{ci}$--$f_{\rm kaw,max}$), ensuring that $\mathcal{R}$ reflects KAW-band power only. Throughout this analysis, $\delta B_\perp$, $\delta B_\parallel$, $\delta E_\perp$, and $\delta E_\parallel$ denote the perpendicular and parallel parts of the measured field fluctuations, taken relative to the local field direction $\hat{\mathbf{b}}(t) = \mathbf{B}(t)/|\mathbf{B}(t)|$ obtained from the FGM field. They are fluctuation components: the background is removed by the band restriction itself rather than by a separate running-mean subtraction, since bandpass-filtering to the KAW range eliminates the DC and slowly varying field (and, in the frequency domain, the steady field contributes only at $f=0$, outside the KAW band). For the time-domain ratio $\mathcal{R}$, $|\delta E_\perp|$ and $|\delta B_\perp|$ are the perpendicular amplitudes after bandpass-filtering to the KAW range ($f_{ci}$--$f_{\rm kaw,max}$); the frequency-domain compressibility $C_\parallel$ uses the same field-aligned decomposition resolved across frequency. The $E_\parallel$ projection in subsequent sections uses the same $\hat{\mathbf{b}}(t)$. As shown in Figure~\ref{fig:overview}(d), the measured ratio during the analysis interval is $\mathcal{R} = 2.5 \pm 1.2$, consistently exceeding the MHD limit throughout the observation period. The scatter in this ratio reflects the turbulent, intermittent nature of the fluctuations rather than instrumental noise. This elevated ratio provides strong evidence that the observed turbulence is mediated by kinetic Alfv{\'e}n waves rather than MHD modes.

To further quantify the magnetic anisotropy characteristic of KAWs, we compute the magnetic compressibility as a function of frequency:
\begin{equation}
C_{\parallel} = \frac{|\delta B_{\parallel}|^2}{|\delta B_{total}|^2} = \frac{|\delta B_{\parallel}|^2}{|\delta B_{\parallel}|^2 + |\delta B_{\perp}|^2}.
\label{eq:compressibility}
\end{equation}
This quantity measures the fraction of magnetic fluctuation power in the field-aligned (compressive) component. For ideal MHD Alfv{\'e}n waves, $C_{\parallel} = 0$ since the fluctuations are purely transverse. Isotropic turbulence yields $C_{\parallel} = 1/3$ by symmetry (equal power in all three field components \citep{gary1986,gary2009short}), while compressive modes such as fast magnetosonic waves exhibit $C_{\parallel} \sim 1$. KAWs are characterized by $C_{\parallel} \lesssim 0.1$ at kinetic scales by linear theory \citep{stasiewicz2000,salem2012identification} and confirmed in solar wind and magnetosheath observations \citep{sahraoui2009, sahraoui2012}, reflecting their predominantly transverse magnetic perturbations.

\begin{figure}[htbp]
\centering
\includegraphics[width=0.80\linewidth]{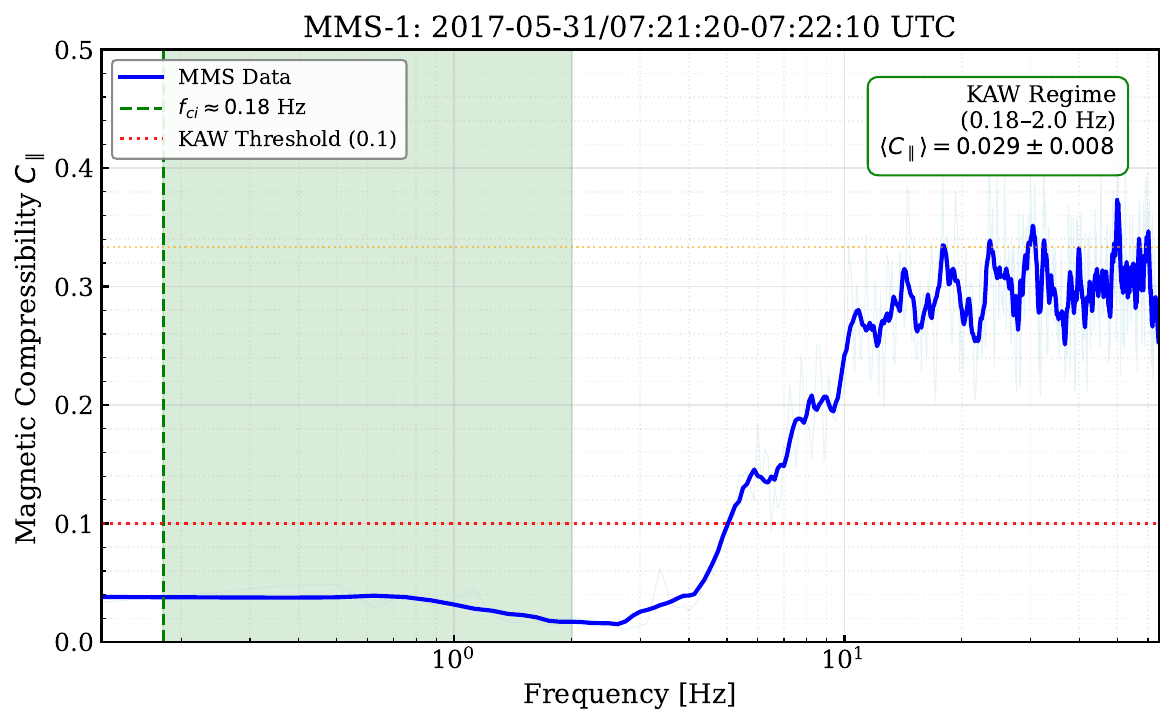}
\caption{Frequency dependence of magnetic compressibility, defined as $C_{\parallel} = |\delta B_{\parallel}|^2 / |\delta B_{total}|^2$. The blue trace shows the smoothed MMS data. The vertical green dashed line marks the local ion cyclotron frequency ($f_{ci} \approx 0.18$~Hz). The horizontal red dotted line indicates the empirical upper limit for KAW identification ($C_{\parallel} \approx 0.1$). The green shaded region highlights the kinetic range ($0.18$--$2.0$~Hz), where the observed mean compressibility is $\langle C_{\parallel} \rangle = 0.029 \pm 0.008$. This value lies well below the $0.1$ threshold, confirming the predominantly transverse nature of the magnetic fluctuations.}
\label{fig:compressibility}
\end{figure}

Figure~\ref{fig:compressibility} presents the magnetic compressibility as a function of frequency for the analysis interval. The data reveal that the compressibility remains consistently low across the transition from MHD to kinetic scales. Within the identified KAW regime ($0.18$--$2.0$~Hz), the mean compressibility is $\langle C_{\parallel} \rangle \approx 0.029 \pm 0.008$. This value is well below both the empirical KAW threshold of $0.1$ and the isotropic limit of $1/3$. This persistent magnetic anisotropy ($|\delta B_{\perp}|^2 \gg |\delta B_{\parallel}|^2$) provides robust confirmation that the turbulence is mediated by shear Alfv{\'e}nic (i.e., transversely polarized) fluctuations rather than compressive magnetosonic modes.

\subsection{Power Spectral Density}
\label{subsec:psd_break}

\begin{figure}[htbp]
\centering
\includegraphics[width=0.80\linewidth]{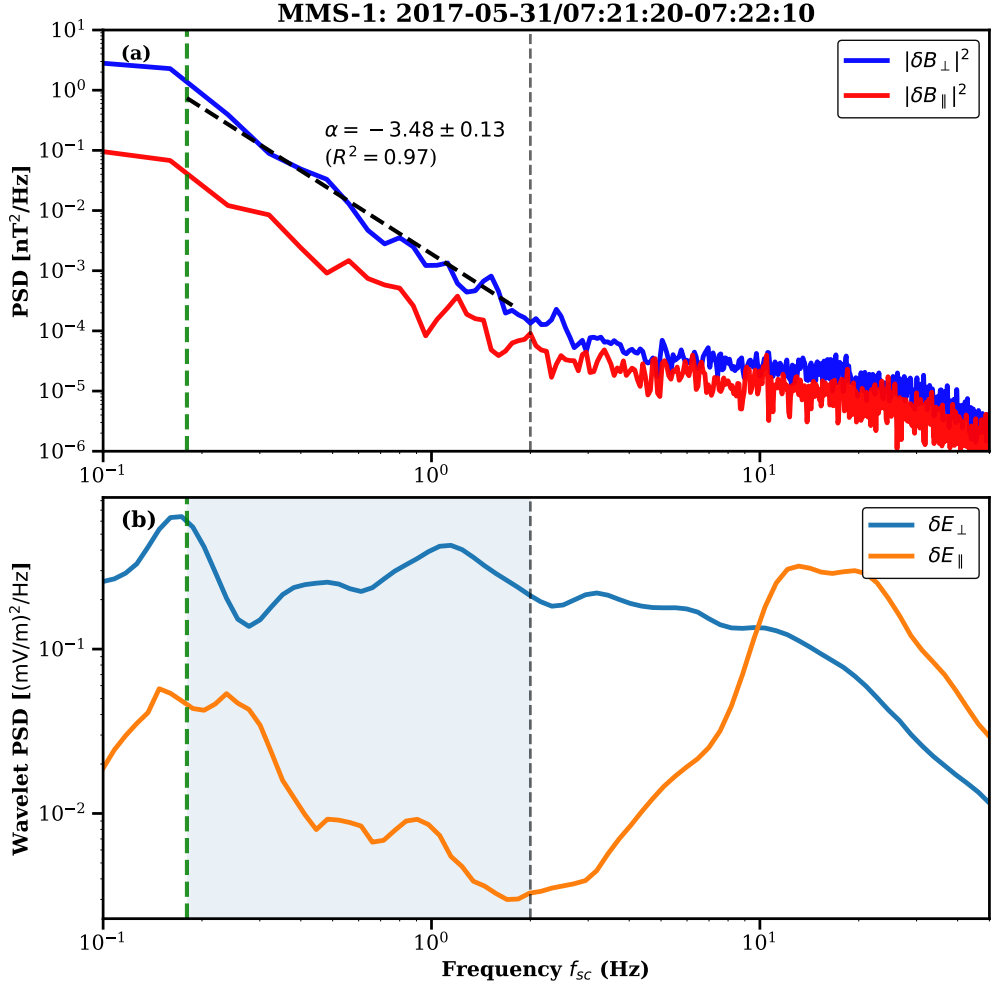}
\caption{Two-panel power spectral density summary of field-aligned fluctuations for MMS-1 during 2017-05-31/07:21:20--07:22:10~UTC. \textbf{(a)} Magnetic-field PSDs in the field-aligned coordinate system: perpendicular $|\delta B_{\perp}|^2$ (blue) and parallel $|\delta B_{\parallel}|^2$ (red). The vertical green dashed line marks the local ion cyclotron frequency ($f_{ci}\approx0.18$~Hz). The black dashed line denotes a power-law fit to the perpendicular magnetic spectrum in the kinetic range, yielding $\alpha=-3.48\pm0.13$ with $R^2=0.97$. The vertical gray dashed line marks the adopted upper bound of the KAW band ($f_{\mathrm{kaw,max}}=2$~Hz). \textbf{(b)} Electric-field spectra for $\delta E_{\perp}$ and $\delta E_{\parallel}$ obtained from a Morlet-wavelet time--frequency analysis and averaged over the same interval. The shaded region highlights the KAW band between $f_{ci}$ and $f_{\mathrm{kaw,max}}$.}
\label{fig:psd}
\end{figure}

Power spectral densities (PSDs) of magnetic and electric field fluctuations are computed using two complementary approaches. For stationary spectral characterization, we employ Welch's method with Hanning windows and 50\% overlap, yielding a frequency resolution of 0.05~Hz over the analysis interval. To resolve transient features in the time--frequency domain, we apply continuous wavelet transforms using the Morlet wavelet with nondimensional frequency $\omega_0 = 6$, providing a balance between temporal and spectral localization \citep{torrence1998}. In practice, Welch's method is applied to the magnetic-field PSDs (the compressibility in Figure~\ref{fig:compressibility} and the kinetic-range slope in Figure~\ref{fig:psd}a), whereas the Morlet-wavelet time--frequency decomposition is used for the electric-field spectra (Figure~\ref{fig:psd}b). Spectral slopes in the kinetic range are determined by least-squares fitting of $\log(\mathrm{PSD})$ versus $\log(f)$ for the perpendicular magnetic spectrum over the interval $f_{ci} < f \le f_{\mathrm{kaw,max}}$ (here $f_{\mathrm{kaw,max}}=2$~Hz). The upper boundary $f_{\mathrm{kaw,max}}=2$~Hz is set at the frequency above which the magnetic compressibility $C_\parallel$ begins to rise above the KAW threshold of $\sim$0.1 (Figure~\ref{fig:compressibility}), marking the transition to higher-frequency compressive or noise-dominated contributions; the spectral fit is restricted to frequencies where the KAW mode identification is unambiguous. Uncertainties on spectral indices are taken from the regression standard error.

We note that the observed bulk ion velocity ($|\mathbf{V}_i| \approx 150$~km/s, $|\mathbf{V}_i|/v_A \approx 0.23$) does not satisfy the standard Taylor hypothesis condition $|\mathbf{V}_{\mathrm{flow}}| \gg v_A$. The implications for spectral interpretation are discussed in Section~\ref{subsec:limitations}.

Figure~\ref{fig:psd} summarizes the spectral properties of the field-aligned fluctuations. Panel~(a) shows the magnetic-field power spectral densities in the field-aligned coordinate system, where the blue and red traces represent the perpendicular ($|\delta B_{\perp}|^2$) and parallel ($|\delta B_{\parallel}|^2$) components, respectively. A clear spectral break occurs near $f \approx 0.18$~Hz, close to the local ion cyclotron frequency ($f_{ci}$; green dashed line), separating the MHD-range behavior from the sub-ion kinetic regime. For $f>f_{ci}$, the perpendicular power dominates the parallel component ($|\delta B_{\perp}|^2 \gg |\delta B_{\parallel}|^2$), demonstrating a persistent anisotropy consistent with Alfv{\'e}nic/KAW turbulence and inconsistent with predominantly compressive magnetosonic fluctuations \citep{wygant2002, chaston2008, sahraoui2009}.

Panel~(b) presents the electric-field spectra for the perpendicular and parallel components, $\delta E_{\perp}$ and $\delta E_{\parallel}$, obtained from a Morlet-wavelet time--frequency decomposition and averaged over the analysis interval. The KAW band used throughout this study is highlighted between $f_{ci}$ and $f_{\mathrm{kaw,max}}=2$~Hz (shaded region). In this band, $\delta E_{\perp}$ provides the dominant contribution while a finite $\delta E_{\parallel}$ is present, consistent with kinetic Alfv{\'e}n wave electrodynamics and the occurrence of parallel electric-field structures discussed in Section~\ref{subsec:parallel_electric}. At frequencies above $f_{\mathrm{kaw,max}}$, the electric-field spectra should be interpreted cautiously because projection effects, residual spin-related contamination, and increasing measurement noise can influence the apparent $\delta E_{\parallel}$ level. In the frequency range much above the KAWs, the PSD of $\delta E_{\parallel}$ is enhanced and exceeds that of $\delta E_{\perp}$ whereas PSD of $\delta B$ remains at low level. This frequency range corresponds to that of electrostatic ion acoustic waves \citep{zhang2022observations}.

To characterize the kinetic-range cascade, we fit a power law to the \emph{perpendicular magnetic} spectrum above $f_{ci}$, restricting the fit to the KAW band used in this study ($f_{ci} < f \leq f_{\mathrm{kaw,max}}=2$~Hz). The regression yields a spectral slope of $\alpha=-3.48\pm0.13$ with $R^2=0.97$, indicating that a single power law provides a strong description of the kinetic-range scaling over this interval. This scaling is significantly steeper than the standard Kolmogorov index ($-5/3$) characteristic of the MHD inertial range and also steeper than theoretical predictions for undamped KAW turbulence, which typically range from $-7/3 \approx -2.33$ \citep{boldyrev2012} to $-8/3 \approx -2.67$ \citep{howes2008}.

The steeper-than-predicted kinetic-range slope observed here is consistent with enhanced dissipation at sub-ion scales, as has been reported in other magnetospheric environments \citep{sahraoui2009, huang2014, roberts2017} and is consistent with numerical predictions showing spectral steepening when damping is included \citep{chettri2025damped}. Possible mechanisms contributing to this steepening include: (1) electron Landau damping of KAWs, which removes energy from the turbulent cascade at kinetic scales; (2) intermittency, wherein energy concentrates in localized coherent structures; and (3) enhanced dissipation in ion-scale current sheets. The combination of strong perpendicular anisotropy, the spectral break near $f_{ci}$, and the steep kinetic-range slope supports an interpretation of KAW turbulence undergoing significant damping. Due to the limited event duration, we do not attempt a robust characterization of the low-frequency MHD-range slope ($f<f_{ci}$).

\subsection{Parallel Electric Fields, Density Structures, and Correlation Analysis}
\label{subsec:parallel_electric}

\begin{figure}[htbp]
\centering
\includegraphics[width=0.85\linewidth]{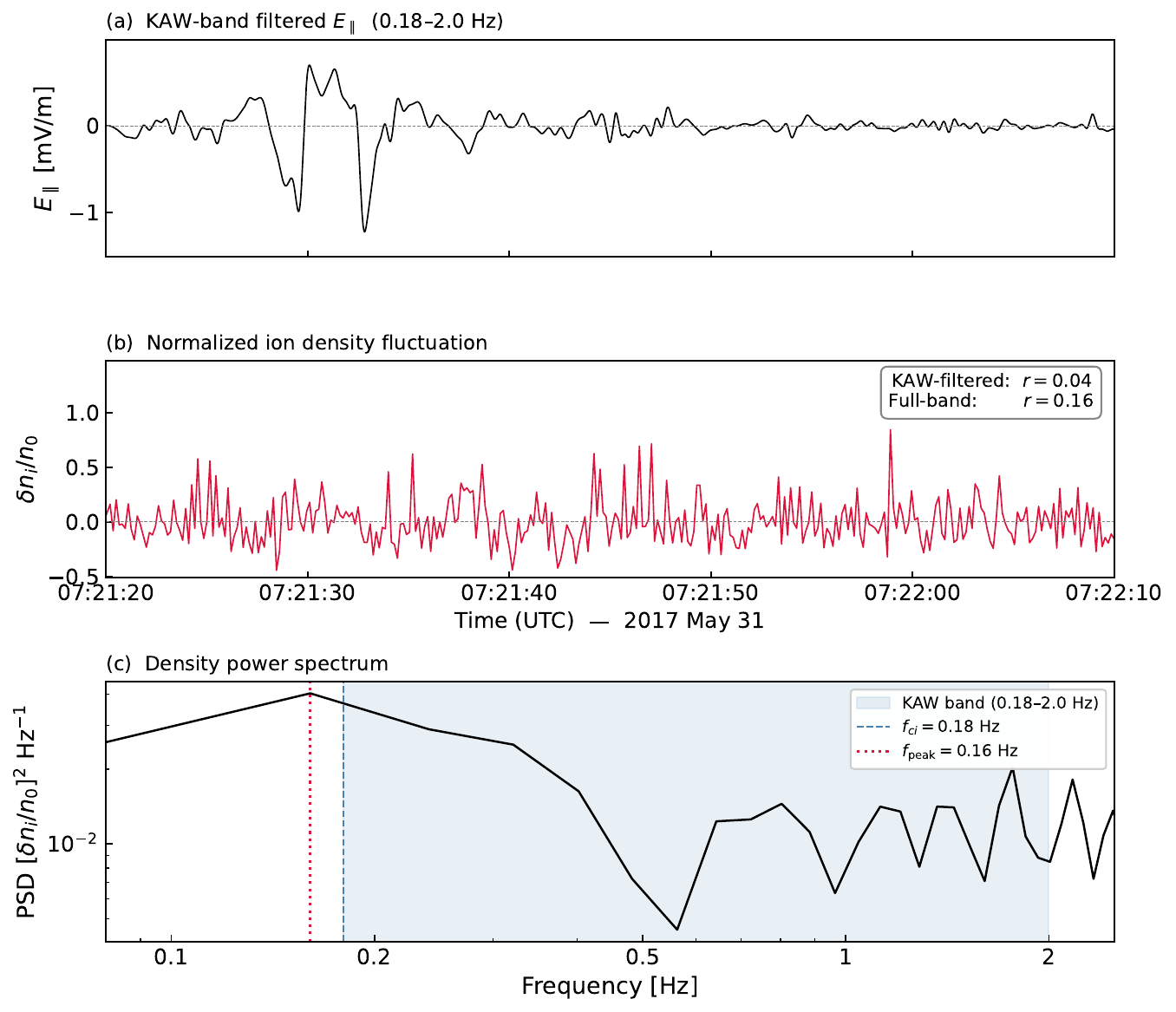}
\caption{KAW-band filtered parallel electric field, ion density fluctuation, and ion density power spectrum during the analysis interval. \textbf{(a)} $E_{\parallel}$ bandpass-filtered to the KAW range (0.18--2.0~Hz; 4th-order Butterworth, zero-phase), computed from $\mathbf{E}\cdot\hat{\mathbf{b}}$ on the electric-field timeline. \textbf{(b)} Normalized ion density fluctuation $\delta n_i/n_0$ (MMS/FPI DIS), with $n_0$ defined as an $8$-s moving-average background; large-amplitude perturbations ($\delta n_i/n_0 \sim \pm 0.5$--$0.6$) co-occur with the wave-active interval. A zero-lag Pearson correlation over 07:21:30--07:21:45~UTC between the KAW-filtered $E_\parallel$ and similarly filtered $\delta n_i/n_0$ gives $r=0.04$, indicating no significant linear association within the KAW band. \textbf{(c)} Power spectral density (PSD) of $\delta n_i/n_0$ (Welch method, Hann window) over the full analysis interval. The blue shaded region marks the KAW band ($f_{ci}$--2.0~Hz). The density PSD peaks at $f_{\rm peak}\approx 0.16$~Hz (dotted red line), just below $f_{ci}=0.18$~Hz, indicating that ion density fluctuations are dominated by sub-ion-cyclotron dynamics rather than KAW-frequency compressive coupling.}
\label{fig:correlation}
\end{figure}

Parallel electric fields provide a direct diagnostic of non-ideal plasma dynamics and are therefore central to assessing how electromagnetic fluctuations exchange energy with particles. In an ideal Alfv{\'e}nic response one expects $E_{\parallel}\approx 0$, whereas finite $E_{\parallel}$ arises naturally for KAWs \citep{kletzing1994electron,wygant2002, bian2010parallel} and, more generally, whenever dispersive and charge-separation effects become important \citep{wygant2002,tian2026evidence}. Density structuring quantifies the compressive response of the plasma and may reflect either wave polarization (including KAW-associated compressibility) or independent boundary-layer and large-scale dynamics. Figure~\ref{fig:correlation} is therefore used to address a specific question relevant to interpreting the interval: \emph{do the most intense $E_{\parallel}$ bursts show a simple linear correspondence with the contemporaneous density perturbations?}

The parallel electric field is constructed on the electric-field timeline as
\begin{equation}
E_{\parallel}(t)=\mathbf{E}(t)\cdot\hat{\mathbf{b}}(t) =\frac{\mathbf{E}(t)\cdot\mathbf{B}(t)}{|\mathbf{B}(t)|},
\end{equation}
where $\mathbf{B}$ is linearly interpolated to the higher-cadence electric-field timestamps before projection. The normalized ion density fluctuation is defined as
\begin{equation}
\frac{\delta n_i(t)}{n_0(t)}=\frac{n_i(t)-n_0(t)}{n_0(t)},
\end{equation}
with the background $n_0(t)$ obtained from an $8$-s moving average of $n_i(t)$ (MMS/FPI DIS ion moments; $n_i \approx n_e$ by quasi-neutrality at KAW frequencies). This choice suppresses slow drifts and highlights density structure on timescales shorter than the smoothing window. For plotting, the density series is displayed on the same time axis as $E_{\parallel}$ by interpolation; this is done to enable visual comparison and does not introduce additional information beyond the native moment cadence. To probe coupling within the KAW frequency band specifically, $E_{\parallel}$ is additionally bandpass-filtered (4th-order Butterworth, zero-phase) to 0.18--2.0~Hz, and $\delta n_i/n_0$ is similarly filtered to the same band prior to computing the correlation.

Figure~\ref{fig:correlation} shows the KAW-band filtered $E_{\parallel}$ (top panel), which retains fluctuation power in the 0.18--2.0~Hz range while suppressing the unresolved low-frequency background. Over the same interval the ion density exhibits large fractional variations, with $\delta n_i/n_0$ reaching values of order $\sim +0.6$ during enhancements and $\sim -0.5$ during depletions (bottom panel). The temporal overlap of intense $E_{\parallel}$ bursts with strong density structuring indicates that non-ideal fields are present while the plasma is also highly inhomogeneous; however, overlap alone does not imply that the two signals are dynamically locked.

To quantify their relationship within the plotted processing chain, a zero-lag Pearson correlation coefficient ($r$) is computed over 07:21:30--07:21:45~UTC using the KAW-band filtered $E_{\parallel}(t)$ and the similarly filtered $\delta n_i/n_0(t)$ placed on the same time grid. The resulting value, $r=0.04$, indicates that within this window there is no evidence for a simple linear, zero-lag association between the instantaneous signed $E_{\parallel}$ waveform and the background-removed density fluctuation defined by the adopted $n_0(t)$. The density power spectral density (PSD) peaks at $f_{\rm peak} \approx 0.16$~Hz, just below the KAW band lower bound ($f_{ci} = 0.18$~Hz), indicating that the dominant density variability is driven by sub-ion-cyclotron dynamics rather than intra-KAW-band compressive coupling. The 0.10--2.0-Hz comparison gives a weak correlation ($r=0.23$), and the 30-ms FPI Dual Electron Spectrometer (DES) partial electron density above 42.95~eV shows the same broad 0.1--0.2-Hz maximum as the DIS ion density. The 8-second smoothing window yields a low-pass cutoff of approximately $1/(8\,\text{s}) \approx 0.125$~Hz, well below $f_{ci} = 0.18$~Hz, ensuring that $n_0(t)$ captures sub-KAW-frequency plasma gradients while retaining KAW-band variability in $\delta n_i/n_0$. The $\delta E_{\parallel}$ spectrum (Figure~\ref{fig:psd}b) exhibits peak power near $f_{ci}$, as expected for KAWs whose finite $E_{\parallel}$ arises through electron pressure gradients and finite gyroradius effects at $k_\perp\rho_i \sim 1$; the slight offset between this $E_{\parallel}$ spectral peak and the density PSD peak at $f_{\rm peak} \approx 0.16$~Hz contributes to the near-zero KAW-band correlation. This argues against an interpretation in which the largest $E_{\parallel}$ spikes are merely proportional to the large-scale density enhancement/depletion captured by the $8$-s background. Instead, the intermittency in $E_{\parallel}$ is consistent with localized non-ideal structure and/or dispersive dynamics that need not track the moment-scale density variations.

It is emphasized that $r\simeq 0$ is a statement about \emph{linear, zero-lag} association under the specific filtering and time-base choices, and it does not exclude time-lagged coupling, nonlinear dependence, or scale-dependent relationships. The consistency of the result across both the full-band ($r=0.16$) and KAW-filtered ($r=0.04$) analyses confirms that the absence of linear coupling is not an artefact of broadband contamination, and supports the conservative conclusion that the observed parallel-field bursts are not straightforwardly explained by the large-amplitude density structuring alone.

\subsection{Energy Conversion Proxies in the Electron Frame}
\label{subsec:energy_conversion}

To complement the spectral and correlation diagnostics, a cadence-matched proxy for local field--particle energy exchange is evaluated using the work density measures $J_{\parallel}E_{\parallel}$ and $\mathbf{J}\cdot\mathbf{E}'$. We note that $\mathbf{J}$ and $\mathbf{E}$ here are total-field quantities, not fluctuation components; the proper wave-particle energy transfer diagnostic is $\delta\mathbf{J}\cdot\delta\mathbf{E}$ (fluctuating current and electric field; see e.g.\ \citealt{chen2019, liu2025}), requires isolating the fluctuating currents and fields against a well-defined background, which necessitates a statistically stationary interval. Because the background cannot be robustly defined during this brief ($\sim 50\text{ s}$), non-stationary boundary-layer crossing, a reliable $\delta \mathbf{J} \cdot \delta \mathbf{E}$ cannot be obtained. Consequently, we utilize $J_{\parallel} E_{\parallel}$ and $\mathbf{J} \cdot \mathbf{E}'$ as conservative energy-conversion proxies to indicate when and where field--plasma coupling is enhanced, rather than as direct wave-particle energy transfer rates. Here,
\begin{equation}
\mathbf{J}(t)= e\,n_e(t)\big[\mathbf{V}_i(t)-\mathbf{V}_e(t)\big],
\end{equation}
is obtained from MMS/FPI ion and electron bulk moments, and the electron-frame electric field is
\begin{equation}
\mathbf{E}'(t)=\mathbf{E}(t)+\mathbf{V}_e(t)\times\mathbf{B}(t).
\end{equation}
All quantities are evaluated on the ion-moment time base: $\mathbf{E}$ (EDP) and $\mathbf{B}$ (FGM) are linearly interpolated to the ion-moment timestamps prior to computing $\hat{\mathbf{b}}=\mathbf{B}/|\mathbf{B}|$, $E_{\parallel}=\mathbf{E}\cdot\hat{\mathbf{b}}$, and $J_{\parallel}=\mathbf{J}\cdot\hat{\mathbf{b}}$. This cadence-matching is required for a physically consistent product and naturally suppresses the narrowest $E_{\parallel}$ spikes that are visible at the native EDP sampling in the overview figure.

\begin{figure}[htbp]
\centering
\includegraphics[width=0.85\linewidth]{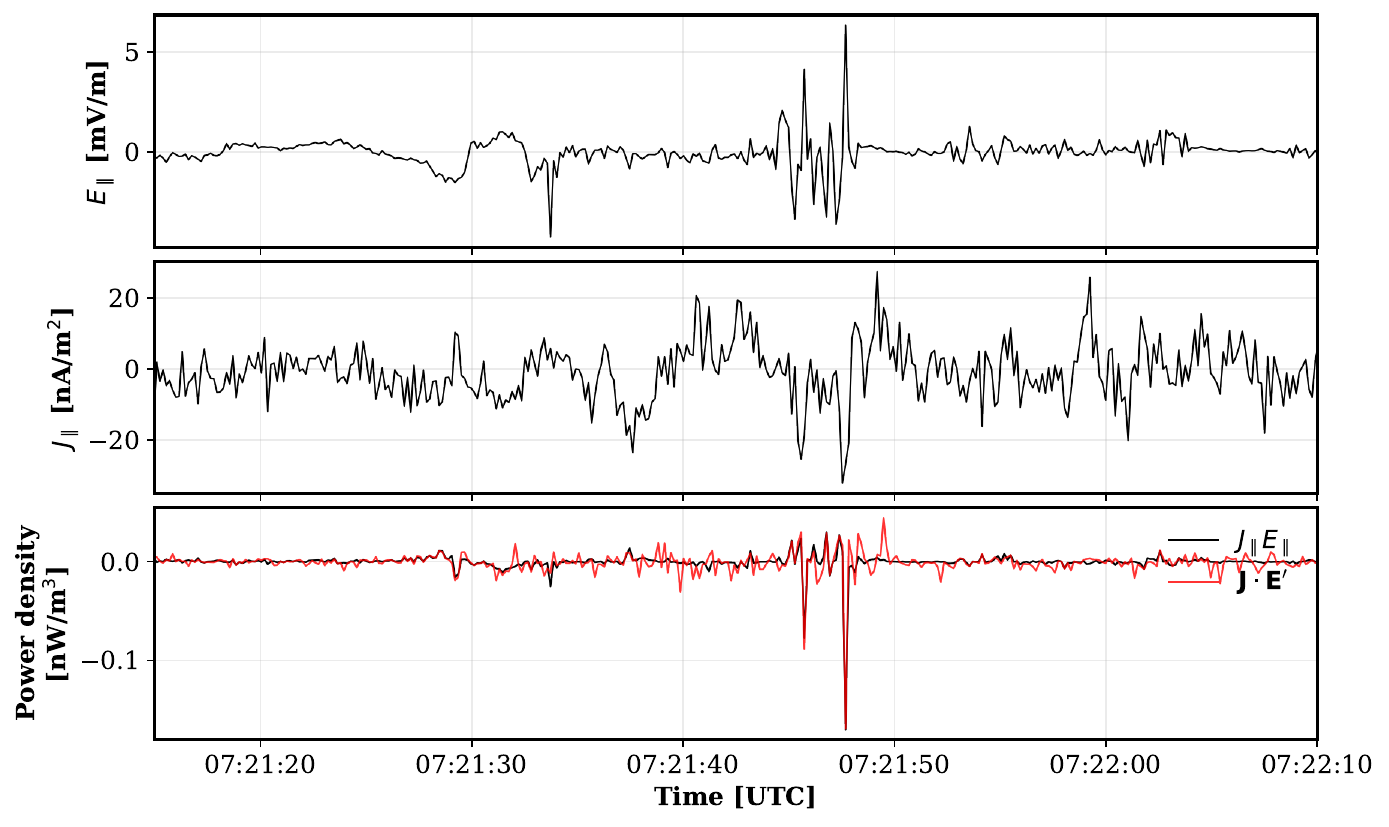}
\caption{Cadence-matched energy-conversion proxies during 2017-05-31/07:21:15--07:22:10~UTC. Top: signed $E_{\parallel}=\mathbf{E}\cdot\hat{\mathbf{b}}$ evaluated on the ion-moment time base after interpolating EDP $\mathbf{E}$ and FGM $\mathbf{B}$ to the moment timestamps. Middle: $J_{\parallel}$ from moments, using $\mathbf{J}=e n_e(\mathbf{V}_i-\mathbf{V}_e)$. Bottom: work density proxies $J_{\parallel}E_{\parallel}$ (black) and $\mathbf{J}\cdot\mathbf{E}'$ (red), where $\mathbf{E}'=\mathbf{E}+\mathbf{V}_e\times\mathbf{B}$. The bursty excursions (both signs) provide a time-domain proxy for intervals of enhanced field--particle energy exchange on the moment cadence.}
\label{fig:energy_conversion}
\end{figure}

Figure~\ref{fig:energy_conversion} summarizes the resulting time series over 07:21:15--07:22:10~UTC. Intermittent $E_{\parallel}$ excursions coincide with enhanced $J_{\parallel}$ variability, and the corresponding work density proxies show bursty excursions of order $10^{-1}$~nW/m$^{3}$. The two measures, $J_{\parallel}E_{\parallel}$ and $\mathbf{J}\cdot\mathbf{E}'$, track each other closely for much of the interval, indicating that on the moment cadence the inferred energy exchange is largely accounted for by the field-aligned contribution. The sign changes in both proxies carry physical meaning: positive excursions ($J_\parallel E_\parallel > 0$, $\mathbf{J}\cdot\mathbf{E}' > 0$) indicate net field-to-particle energy transfer on the moment cadence, consistent with local particle heating or acceleration; negative excursions indicate the reverse: the plasma drives the electromagnetic field, as expected during wave growth or back-reaction phases in turbulent dissipation. Both signs are anticipated in intermittent KAW turbulence, and the estimate represents local energy exchange rather than a guaranteed net heating rate; it is therefore used here as a conservative indicator of when and where strong field--particle coupling is occurring during the wave-active interval.

\section{Discussion}
\label{sec:discussion}

\subsection{Evidence for KAW Turbulence in the Outer PSBL}

The observations presented here provide multiple lines of evidence that KAW turbulence is active in the outer PSBL/lobe boundary region. First, the electric-to-magnetic field ratio $\mathcal{R} = 2.5 \pm 1.2$ significantly exceeds the MHD Alfv{\'e}n wave value of unity, consistent with the theoretical expectation for KAWs at $k_{\perp}\rho_i \gtrsim 1$ \citep{stasiewicz2000, chaston2008}. Second, the magnetic fluctuations exhibit strong perpendicular anisotropy ($|\delta B_{\perp}|^2 \gg |\delta B_{\parallel}|^2$), a defining characteristic of KAWs that distinguishes them from fast magnetosonic modes or mirror mode structures \citep{gary1986, sahraoui2012}. Third, the spectral break near the ion cyclotron frequency marks the transition from MHD to kinetic physics, as expected when the turbulent cascade reaches ion gyroscale wavelengths. Fourth, the magnetic compressibility $C_{\parallel} \approx 0.03$ in the kinetic range (Figure~\ref{fig:compressibility}) falls well below both the isotropic value ($1/3$) and the empirical KAW threshold (${\sim}0.1$), providing quantitative confirmation that the fluctuations are predominantly transverse. This low compressibility is inconsistent with mirror modes ($C_{\parallel} \sim 0.5$--$1$) or fast magnetosonic turbulence, further supporting the KAW interpretation.

These findings extend previous observations of KAW turbulence in the central PSBL \citep{wygant2002, chaston2008} to the cooler outer PSBL adjacent to the magnetotail lobes. The relatively low ion temperature ($T_i = 0.65 \pm 0.55$~keV) compared to the typical PSBL values of 1--5~keV suggests that our event captures the transition region where lobe plasma mixes with plasma sheet particles. Despite these cooler conditions, KAW turbulence remains active, indicating that kinetic-scale energy dissipation may operate across a broader range of magnetotail environments than previously documented.

\subsection{Parallel Electric Field Structures and Identification of Dissipation Pathway}

The overview time series shows brief, high-cadence enhancements in $|\delta E_{\parallel}|$ approaching 13--15~mV/m (Figure~\ref{fig:overview}c). In the cadence-matched energy-conversion analysis, $E_{\parallel}$ is evaluated on the ion-moment time base, so the narrowest spikes visible at the native EDP cadence are strongly suppressed (Figure~\ref{fig:energy_conversion}). The strongly intermittent morphology, with characteristic durations well below the MMS spin period ($\sim$20~s), is consistent with localized non-ideal structures (e.g., solitary electric-field structures or phase-steepened packets) embedded within the KAW turbulence, rather than with a purely sinusoidal linear KAW waveform.

Parallel electric fields of this magnitude have the potential to accelerate electrons along the magnetic field direction. The $E_{\parallel}$ largely comes from high frequency fluctuations, which are expected to have rather small wavelengths. Therefore, making a simple $\Delta W \sim eE_\parallel L$ estimate is not appropriate without knowledge of the parallel coherence scale from multi-spacecraft timing. Direct verification requires electron velocity distribution analysis; \citet{zhang2022observations} performed such analysis for this same event, demonstrating electron acceleration via pitch-angle distributions and parallel potential-drop estimates, providing kinematic evidence for the accelerated particle population. The present study complements that work by providing a dynamical diagnostic: the cadence-matched $J_\parallel E_\parallel$ and $\mathbf{J}\cdot\mathbf{E}'$ proxies (Section~\ref{subsec:energy_conversion}) quantify the rate of field--plasma energy exchange, rather than the resulting particle distributions. These are distinct and complementary observational approaches.

The near-zero Pearson correlation ($r$) between the KAW-band filtered $E_{\parallel}$ waveform and the background-removed ion density fluctuation ($r=0.04$, within 0.18--2.0~Hz) indicates that, under the adopted filtering and time-base choices, there is no evidence for a simple linear, zero-lag correspondence between these two quantities. This result does not exclude time-lagged or nonlinear coupling, nor does it rule out scale-dependent relationships, but it argues against an interpretation in which the largest $E_{\parallel}$ spikes are merely proportional to the moment-scale density enhancement/depletion captured by the $8$-s background. Taken together with the bursty $J_{\parallel}E_{\parallel}$ and $\mathbf{J}\cdot\mathbf{E}'$ signatures, the observations are consistent with localized wave--particle energy exchange occurring within a strongly structured boundary-layer plasma. A specific damping channel (e.g., electron Landau damping) is plausible but is not uniquely constrained by the present field-moment diagnostics.

\subsection{Steep Spectral Slope and Enhanced Dissipation}

The measured kinetic range power density spectral slope of $\alpha = -3.48 \pm 0.13$ is steeper than theoretical predictions for undamped KAW turbulence. The observed slope is similar to the $\alpha \approx -3$ to $-4$ values reported in other magnetospheric and solar wind KAW studies where strong damping is inferred \citep{sahraoui2009, alexandrova2012, huang2014}, and matches the $-3$ to $-4$ dissipation-range indices from nonlinear KAW--ion-acoustic Zakharov simulations \citep{chettri2024nonlinear}, supporting efficient kinetic-scale turbulent dissipation. Several factors may contribute to this steepening:

\textit{Landau damping:} In a plasma with $T_i/T_e \approx 2.4$, electron Landau damping of KAWs is expected to be significant at sub-ion scales where $k_{\parallel}v_{th,e} \sim \omega$ \citep{howes2006, schekochihin2009}. The damping rate increases with $k_{\perp}\rho_i$, progressively removing energy from the cascade and steepening the spectrum beyond passive cascade predictions. This effect has been demonstrated numerically by \citet{chettri2025damped}, who showed that including phenomenological Landau damping in a modified nonlinear Schr\"odinger equation steepens the sub-ion range spectrum from $k_\perp^{-8/3}$ to $k_\perp^{-11/3}$ in magnetosheath-like conditions.

\textit{Ion Landau damping of compressive fluctuations:} The relatively high $T_i/T_e$ ratio suppresses freely propagating ion acoustic modes, but in a strongly structured boundary-layer plasma, compressive responses can still be continuously driven by finite-amplitude Alfv{\'e}nic/KAW fluctuations. The co-occurrence of large density variability with the wave-active interval (Figure~\ref{fig:correlation}) is compatible with such coupling, although the present correlation test does not show a simple linear, zero-lag correspondence between the instantaneous signed $E_{\parallel}$ waveform and the background-removed density fluctuation. In this case, any compressive energy injected into ion-acoustic-like responses would be expected to be strongly damped, providing an additional route for energy removal that is not captured in idealized undamped KAW cascade theories \citep{terradas2022, verscharen2024}.

\textit{Intermittency:} Turbulent dissipation concentrated in localized current sheets and coherent structures produces steeper spectral slopes than homogeneous cascade models \citep{osman2011, greco2012}. The impulsive nature of the $E_{\parallel}$ signatures in Figure~\ref{fig:overview}(c) and the bursty character of the energy-conversion proxies (Figure~\ref{fig:energy_conversion}) are both consistent with intermittent energy dissipation.

\subsection{Limitations and Caveats}
\label{subsec:limitations}

\textit{Single event:} This study analyzes a single 50-second interval during one PSBL crossing. Statistical studies across multiple events are needed to assess how representative these observations are of outer PSBL conditions generally.

\textit{Short interval duration:} The 50-second analysis window limits the low-frequency extent of reliable spectral estimates. We therefore do not characterize the MHD inertial range slope or the complete turbulent cascade from injection to dissipation scales.

\textit{Validity of the Taylor hypothesis:} The observed bulk ion velocity ($|\mathbf{V}_i| \approx 150$~km/s, corresponding to $|\mathbf{V}_i|/v_A \approx 0.23$) does not satisfy the standard condition $|\mathbf{V}_{\mathrm{flow}}| \gg v_A$ required for a strict Taylor hypothesis mapping between temporal frequency and spatial wavenumber. However, at kinetic scales ($k_{\perp}\rho_i \gtrsim 1$), the dispersive nature of kinetic Alfv{\'e}n waves reduces the phase speed relative to $v_A$, which can improve the applicability of the hypothesis at the smallest scales considered \citep{howes2014}. Even with this mitigation, the low flow speed introduces systematic uncertainties: spectral indices may be biased by $\Delta\alpha \approx \pm 0.3$--$0.5$, and wavenumber estimates by a factor of $\sim 2$ \citep{klein2014}. At sub-ion scales, kinetic Alfv{\'e}n wave dispersion and anisotropy further complicate a direct frequency--wavenumber interpretation, so the reported kinetic-range slope should be regarded as a frequency-domain characterization of the scaling. To quantify the Doppler shift at the ion gyroscale, we estimate $\Delta f = k_\perp |\mathbf{V}_i|/(2\pi)$ with $k_\perp \approx \rho_i^{-1} \approx 3.1\times10^{-3}$~km$^{-1}$ and $|\mathbf{V}_i| \approx 150$~km/s, giving $\Delta f \approx 0.075$~Hz. This is smaller than $f_{ci} = 0.18$~Hz and much smaller than the KAW band width ($\sim$1.8~Hz), so the spectral break identification and the KAW-band boundaries are only weakly affected. The mode identification via $C_\parallel$ and $\mathcal{R}$ does not require frequency-to-wavenumber conversion and is therefore unaffected by the Doppler correction. Consequently, spectral indices are interpreted with appropriate systematic uncertainty.

\textit{Parallel-field projection uncertainty:} $E_\parallel$ is obtained by projecting $\mathbf{E}$ onto $\hat{\mathbf{b}}$, so the much larger perpendicular field can leak into the parallel channel through the angular uncertainty in $\hat{\mathbf{b}}$, giving a spurious contribution of order $|\delta E_\perp|\sin\Delta\theta$. For a conservative few-degree $\Delta\theta$ and $|\delta E_\perp|$ up to ${\sim}10$~mV/m, this leakage is $\lesssim 0.5$~mV/m, well below the 13--15~mV/m $E_\parallel$ structures, so the strongest parallel-field signatures are robust while the weakest excursions carry larger fractional uncertainty. The moment-derived $J_\parallel E_\parallel$ proxy inherits both this $E_\parallel$ uncertainty and the FPI moment uncertainty in $\mathbf{J}$, and is therefore interpreted as an order-of-magnitude indicator rather than a precise transfer rate.

\textit{Particle distribution analysis:} Direct verification of electron Landau damping would require detailed analysis of electron velocity distributions to identify resonant signatures such as flattening near the parallel phase velocity. While the FPI instrument provides such data, this analysis will be carried out in future work.

\section{Conclusions}
\label{sec:conclusions}

We have presented MMS observations of kinetic Alfv{\'e}n wave turbulence during a PSBL crossing on May~31, 2017. The principal findings are:
\begin{enumerate}
\item \textbf{KAW identification:} The electromagnetic fluctuations exhibit properties consistent with KAW turbulence, including (i) an electric-to-magnetic field ratio $\mathcal{R} = 2.5 \pm 1.2$ exceeding the MHD limit; (ii) strong perpendicular magnetic anisotropy ($|\delta B_{\perp}|^2 \gg |\delta B_{\parallel}|^2$); (iii) low magnetic compressibility in the kinetic range ($C_{\parallel} \approx 0.03$); and (iv) a spectral break near the ion cyclotron frequency ($f_{ci} \approx 0.18$~Hz).

\item \textbf{Steep kinetic-range spectrum:} The perpendicular magnetic spectrum in the kinetic range follows a power law with slope $\alpha = -3.48 \pm 0.13$, steeper than undamped KAW cascade predictions ($-7/3$ to $-8/3$), consistent with substantial energy removal from the cascade at sub-ion scales.

\item \textbf{Parallel fields and density structuring:} Impulsive $E_{\parallel}$ structures reach 13--15~mV/m and large-amplitude ion density fluctuations occur during the wave-active interval ($\delta n/n$ up to ${\sim}68\%$). A zero-lag Pearson correlation computed between the KAW-band filtered (0.18--2.0~Hz) $E_{\parallel}$ waveform and the background-removed ion density fluctuation gives $r=0.04$, indicating no simple linear, zero-lag correspondence under the adopted processing choices. The ion density PSD peaks at $f_{\rm peak}\approx 0.16$~Hz, just below $f_{ci}$, consistent with dominant density variability residing below the KAW band. Together with the bursty $J_{\parallel}E_{\parallel}$ and $\mathbf{J}\cdot\mathbf{E}'$ proxies, these observations are consistent with localized field--particle energy exchange within a strongly structured boundary-layer plasma; a specific damping channel (e.g., Landau damping) is plausible but is not uniquely constrained by the present diagnostics.

\item \textbf{Outer PSBL regime:} The event occurs in a relatively cold, outer PSBL/lobe boundary environment ($T_i \approx 0.65$~keV, $\beta_i \approx 0.29$). The quantitative turbulence diagnostics reported here (kinetic-range spectral slope, magnetic compressibility, KAW-band $E_\parallel$--density coupling, and energy-conversion proxies) provide the characterization of KAW turbulence in the cooler PSBL boundary regime beyond prior identification studies such as \citet{zhang2022observations}, resulting in a more complete picture of kinetic-scale energy conversion at the lobe boundary.
\end{enumerate}

Future work will include particle distribution analyses to test for resonant signatures, multi-spacecraft timing to constrain structure scales and wavevector properties, and surveys across multiple PSBL crossings to assess the prevalence of KAW turbulence and associated kinetic-scale energy conversion in this region.

\section*{Data Availability Statement}

The observational data from the Magnetospheric Multiscale (MMS) mission used in this study are publicly available from the NASA Space Physics Data Facility (\url{https://spdf.gsfc.nasa.gov/pub/data/mms/}). Specifically, we use MMS1 Level-2 burst-mode data products: FGM magnetic field (mms1\_fgm\_brst\_l2), EDP electric field (mms1\_edp\_brst\_l2), and FPI ion/electron moments (mms1\_dis\_brst\_l2, mms1\_des\_brst\_l2) for the interval 2017-05-31 07:21:00--07:22:20~UTC. Data were accessed and processed using the \textsc{PySPEDAS} software package. The analysis scripts used to generate the figures in this work are available from the corresponding author upon reasonable request.

\section*{Declaration of Competing Interest}

The authors declare that they have no known competing financial interests or personal relationships that could have appeared to influence the work reported in this paper.

\section*{Acknowledgments}

The author MKC is grateful to the University Grants Commission (UGC), India, for providing financial support through the Non-NET Fellowship. We acknowledge the MMS mission and instrument teams for providing the data used in this study. We particularly thank the FGM, EDP, and FPI teams for their efforts in producing high-quality burst-mode data products.

\bibliographystyle{apalike}
\bibliography{references}

\end{document}